\documentstyle[twoside]{article}

\catcode`\@=11
\long\def\@makefntext#1{ 
\protect\noindent \hbox to 3.2pt {\hskip-.9pt
$^{{\eightrm\@thefnmark}}$\hfil}#1\hfill} 

\def\thefootnote{\fnsymbol{footnote}}
 \def\@makefnmark{\hbox to 0pt{$^{\@thefnmark}$\hss}}  

\def\ps@myheadings{\let\@mkboth\@gobbletwo
\def\@oddhead{\hbox{} 
\rightmark\hfil\eightrm\thepage}
\def\@oddfoot{}\def\@evenhead{\eightrm\thepage\hfil 
\leftmark\hbox{}}\def\@evenfoot{}
\def\sectionmark##1{}\def\subsectionmark##1{}}

\oddsidemargin=\evensidemargin
\renewcommand{\thefootnote}{\fnsymbol{footnote}}

\newcounter{sectionc}\newcounter{subsectionc}\newcounter{subsubsectionc}
\renewcommand{\section}[1] {\vspace{12pt}\addtocounter{sectionc}{1}
\setcounter{subsectionc}{0}\setcounter{subsubsectionc}{0}\noindent
	{\tenbf\thesectionc. #1}\par\vspace{5pt}}
\renewcommand{\subsection}[1] {\vspace{12pt}\addtocounter{subsectionc}{1}
	\setcounter{subsubsectionc}{0}\noindent
	{\bf\thesectionc.\thesubsectionc. {\kern1pt \bfit #1}}\par\vspace{5pt}}
\renewcommand{\subsubsection}[1] {\vspace{12pt}\addtocounter{subsubsectionc}{1}
	\noindent{\tenrm\thesectionc.\thesubsectionc.\thesubsubsectionc.
	{\kern1pt \tenit #1}}\par\vspace{5pt}}
\newcommand{\nonumsection}[1] {\vspace{12pt}\noindent{\tenbf #1}
	\par\vspace{5pt}}

\newcommand{\textlineskip}{\baselineskip=13pt}
\newcommand{\smalllineskip}{\baselineskip=10pt}

\def\eightcirc{
\begin{picture}(0,0)
\put(4.4,1.8){\circle{6.5}}
\end{picture}}
\def\eightcopyright{\eightcirc\kern2.7pt\hbox{\eightrm c}}

\newcommand{\copyrightheading}[1]
	{\vspace*{-2.5cm}\smalllineskip{\flushleft
	{\eightrm Modern Physics Letters A, #1}\\
	{\eightrm $\eightcopyright$\, World Scientific Publishing
	 Company}\\
	 }}

\newcommand{\publisher}[2]{{\begin{center}\eightrm\smalllineskip
	Received #1\\
	Revised #2
	\end{center}
	}}

\def\abstracts#1#2#3{{
	\centering{\begin{minipage}{4.5in}\baselineskip=10pt\eightrm
	\centerline{ABSTRACT}
	\parindent=0pt #1\par
	\parindent=15pt #2\par
	\parindent=15pt #3
	\end{minipage} }\par}}

\newcommand{\bibit}{\nineit}
\newcommand{\bibbf}{\ninebf}
\renewenvironment{thebibliography}[1]			
	{\ninerm
	 \baselineskip=11pt				
	 \begin{list}{\arabic{enumi}.}			
	{\usecounter{enumi}\setlength{\parsep}{0pt}
	 \setlength{\leftmargin 17pt}{\rightmargin 0pt}	
	 \setlength{\itemsep}{0pt} \settowidth		
	{\labelwidth}{#1.}\sloppy}}{\end{list}}	

\newcounter{itemlistc}
\newcounter{romanlistc}
\newcounter{alphlistc}
\newcounter{arabiclistc}

\newcounter{tempfigtabc}			

\def\pmb#1{\setbox0=\hbox{#1}
	\kern-.025em\copy0\kern-\wd0
	\kern.05em\copy0\kern-\wd0
	\kern-.025em\raise.0433em\box0}

\def\fnm#1{$^{\mbox{\scriptsize #1}}$}
\def\fnt#1#2{\footnotetext{\kern-.3em
	{$^{\mbox{\scriptsize #1}}$}{#2}}}

\def\fpage#1{\begingroup
\voffset=.3in
\thispagestyle{empty}\begin{table}[b]\centerline{\footnotesize #1}
	\end{table}\endgroup}

\def\runninghead#1#2{\pagestyle{myheadings}
\markboth{{\eightit{\quad #1}}\hfill}{\hfill{\eightit{#2\quad}}}}

\font\tenbf=cmbx10
\font\tenit=cmti10
\font\tenit=cmti10
\font\bfit=cmbxti10 at 10pt
 1
 1
 1
\font\ninebf=cmbx9
\font\ninerm=cmr9
\font\nineit=cmti9

\font\eightrm=cmr8
\font\eightit=cmti8

\newcommand{\ffcaption}[1]{
	\setcounter{tempfigtabc}{\thefigure}
	\addtocounter{tempfigtabc}{1}
       {\noindent\parbox{5truein}{\eightrm Fig.~\thetempfigtabc. #1}\par}
       \vskip5mm}

\def\qed{\hbox{${\vcenter{\vbox{                          
   \hrule height 0.4pt\hbox{\vrule width 0.4pt height 6pt
   \kern5pt\vrule width 0.4pt}\hrule height 0.4pt}}}$}}

\runninghead{F. Schlumpf} {The Electric and Magnetic Form Factors of
the neutron}

\begin{document}
\normalsize\textlineskip
\thispagestyle{empty}
\setcounter{page}{1}

\renewcommand{\thefootnote}{\fnsymbol{footnote}} 

\copyrightheading{Vol. 0, No. 0 (1992) 000--000}

\vspace*{0.88truein}

\fpage{1}
\centerline{\bf THE ELECTRIC AND MAGNETIC FORM FACTORS}
\vspace*{0.035truein}
\centerline{\bf OF THE NEUTRON}
\vspace{0.37truein}
\centerline{\footnotesize FELIX SCHLUMPF}
\vspace*{0.015truein}
\centerline{\footnotesize\it Stanford Linear Accelerator Center,
Stanford University}
\baselineskip=10pt
\centerline{\footnotesize\it Stanford, California 94309, USA}
\publisher{(received date)}{(revised date)}

\vspace*{0.21truein}
\abstracts{\noindent We derive the
electric and magnetic form factors of the neutron in the
framework of a relativistic constituent quark model. Our parameter free
prediction agrees well with a recent, accurate measurement. The
relativistic features of the model and the specific form of the wave
function are essential for the result. Comparisons are made to other models
based on VMD, PQCD and QCD sum rules.}{}{}

\setcounter{footnote}{0}
\renewcommand{\thefootnote}{\alph{footnote}}

\vspace*{0.15truein}
\textlineskip
\noindent
A recent measurement $^1$ of the neutron electromagnetic form
factors, $G_{En}(Q^2)$ and $G_{Mn}(Q^2)$, greatly increased the $Q^2$ range
of previous data $^2$ and has significantly smaller errors. For
the first time it is therefore possible to distinguish theoretical models
with respect to experimental data from the neutron form factors. In the low
$Q^2$ region, vector meson dominance (VMD) models $^3$ are
traditionally used to make predictions for the form factor. For sufficiently
high momentum transfer perturbative QCD (PQCD) $^4$ predicts the
$Q^2$ dependence of the form factors. To describe the behavior at
intermediate values of $Q^2$ the parameterization of Ref.~5
uses the VMD form at low $Q^2$, constrained by PQCD results at high $Q^2$.
There are additional models which predict the neutron form factors.
Reference~6 describes a relativistic constituent quark model
which is similar to our approach. QCD sum rules are used in
Ref.~7 to fix the parameters of the soft quark functions for
calculating the form factors. None of these theoretical models are in good
agreements with the data for both form factors.

We recently investigated the predictive power of a relativistic constituent
quark model formulated on the light-front.$^{8,9}$ It
provides a simple model wherein we have overall an excellent and consistent
picture of the magnetic moments and the semileptonic decays of the baryon
octet. The parameters of the model have been fixed in Ref.~9 so
that we have a parameter free prediction of the neutron electromagnetic
form factors.

The light-front dynamic is a convenient scheme for dealing with a
relativistic system.
If we introduce the light-front variables $p^\pm\equiv p^0\pm p^3$,
the Einstein mass relation $p_\mu p^\mu = m^2$ is linear in $p^-$ and
linear in $p^+$, in contrast to the quadratic form in $p^0$ and $\vec p$ in
the usual dynamical scheme. A consequence is a single solution of the mass
shell relation in terms of $p^-$, in contrast to two solutions for $p^0$:
\begin{equation}
p^- = (p_\perp^2 + m^2)/p^+\;,\quad  p^0 = \pm \sqrt{\vec p\,^2 + m^2}\;.
\end{equation}
The quadratic relation of $p^-$ and $p_\perp \equiv (p^1,p^2)$ in the above
Equation resembles the nonrelativistic scheme,$^{10}$ and the
variable $p^+$ plays the role of ``mass'' in this nonrelativistic analogy.
It is therefore a good idea to introduce relative variables like the Jacobi
momenta when dealing with several particles. As in the nonrelativistic
scheme such variables allow us to decouple the center of mass motion from
the internal dynamics. The light-front scheme shows
another attractive feature that it has in common with the infinite momentum
technique.$^{11}$ In terms of the old fashioned perturbation theory,
the diagrams with quarks created out of or annihilated into the vacuum do
not contribute. The usual $qqq$ quark structure is therefore conserved as
in the nonrelativistic theory. It is, however, harder to get the hadron
states to be eigenfunctions of the spin operator.$^{12}$

The light-front formalism is specified by the invariant hypersurface $x^+ =
x^0+x^3 =$ constant. The following notation is used: The four-vector is
given by $x = (x^+,x^-,x_\perp)$, where $x^\pm = x^0 \pm x^3$ and
$x_\perp=(x^1,x^2)$. Light-front vectors are denoted by an arrow $\vec x =
(x^+,x_\perp)$, and they are covariant under kinematic Lorentz
transformations.$^{13}$ The three momenta $\vec p_i$ of the quarks
can be transformed to the total and relative momenta to facilitate the
separation of the center of mass motion:$^{14}$
\begin{eqnarray}
\vec P&=&\vec p_1+\vec p_2+\vec p_3, \nonumber\\
\xi&=&{p_1^+\over p_1^++p_2^+} , \quad
\eta={p_1^++p_2^+\over P^+} ,\nonumber\\
k_\perp&=&(1-\xi)p_{1\perp}-\xi p_{2\perp} ,\nonumber\\
K_\perp&=&(1-\eta)(p_{1\perp}+p_{2\perp})-\eta p_{3\perp} .
\end{eqnarray}
Note that the four-vectors are not conserved, i.e., $p_1+p_2+p_3\not= P$.
In the light-front dynamics the Hamiltonian takes the form
\begin{equation}
H={P^2_\perp + M^2 \over 2P^+} ,
\end{equation}
where $M$ is the mass operator with the interaction term $W$
\begin{eqnarray}
M &=&M_0+W , \nonumber\\
M_0^2&=&{K_\perp^2\over \eta(1-\eta)}+{M_3^2\over \eta}+{m^2\over 1-\eta},
\label{eq:2.3} \nonumber\\
M_3^2&=&{k_\perp^2+m^2\over \xi (1-\xi)} ,
\end{eqnarray}
with $m$ being the mass of the constituent quarks. To get a clearer
picture of $M_0$ we transform to $k_3$ and $K_3$ by
\begin{eqnarray}
\xi&=&{E_1+k_3\over E_1+E_2} , \quad \eta={E_{12}+K_3\over E_{12}+E_3} ,
\nonumber\\
E_{1/2}&=&({\bf k}^2+m_{1/2}^2)^{1/2} ,
E_{3}=({\bf K}^2+m_{3}^2)^{1/2} ,
E_{12}=({\bf K}^2+M_{3}^2)^{1/2} ,
\end{eqnarray}
where ${\bf k}=(k_1,k_2,k_3)$, and ${\bf K}=(K_1,K_2,K_3)$.
The expression for the mass operator is now simply
\begin{equation}
M_0=E_{12}+E_3 ,\quad  M_3=E_1+E_2 .
\end{equation}

The diagrammatic approach to light-front theory is well known.$^{15,16}$
It provides in principal a framework
for a systematic treatment of higher-order gluon exchange. In
this work we limit ourselves to the tree graph. Since we set
$Q^+=0$ we can preserve the correct $qqq$ structure of
the vertex. All relevant matrix elements that we investigate are
related to
\begin{equation}
\left< \vec p\,'\left|\bar q \gamma^+ q\right| \vec p\right>
\sqrt{P^{'+}P^+} \equiv M^+ ,
\end{equation}
where the state $|\vec p\,\rangle\equiv|p\rangle/\sqrt{p^+}$ is
normalized according to
\begin{equation}
\left< \vec p\,' | \vec p \,\right>=\delta (\vec p\,' -\vec p\,) .
\end{equation}
The matrix element $M^+$ can be written in terms of wave functions as:
$^9$
\begin{equation}
M^+=3{N_c\over (2\pi)^6}\int d^3kd^3K\left({E'_3E'_{12}M_0\over
E_3E_{12}M_0'}\right)^{1/2}
\Psi^\dagger({\bf k}',{\bf K}')\Psi({\bf k},{\bf K}) ,
\label{eq:me}
\end{equation}
where $K'_\perp = K_\perp + \eta Q_\perp$, and $N_c$ being the number of
colors.

The electromagnetic current matrix element for the transition
$n \to n'\gamma$ can be written
in terms of two form factors taking into account current and parity
conservation:

\begin{equation}
\left< n',\lambda ' p' \left| J^\mu \right|
n,\lambda p\right> =
\bar u_{\lambda '}(p') \left[ F_1(Q^2)\gamma^\mu + {F_2(Q^2) \over
2 M_n}i\sigma^{\mu\nu}Q_\nu \right] u_\lambda (p)
\label{eq:3.1}
\end{equation}
with momentum transfer $Q = p' - p$, and the current $J^\mu=
e \bar q \gamma^\mu q$.
In order to use Eq.~(\ref{eq:me}) we express the form factors in
terms of the $+$ component of the current:
\begin{eqnarray}
F_1(Q^2) &=&\left< n',\uparrow\left| J^+\right| n,
\uparrow\right> ,\nonumber\\
Q_\perp F_2(Q^2) &=&-{2M_n}\left< n',\uparrow\left|
J^+\right| n,\downarrow\right> .
\end{eqnarray}
For $Q^2 = 0$ the form factors $F_1$ and $F_2$ are respectively equal
to the charge and the anomalous magnetic moment in units $e$ and
$e/M_N$. The Sachs form factors are defined as $G_M=F_1+F_2$ and
$G_E=F_1-\tau F_2$ with $\tau=Q^2/4M_n^2$.

Since the center of mass motion can be separated from the internal
motion, the wave function $\Psi$ is a
function of the relative momenta ${\bf k}$ and ${\bf K}$. The product
$\Psi = \Phi\chi\phi$ with $\Phi =$ flavor, $\chi=$ spin, and $\phi=$
momentum distribution, is a symmetric function. The neutron wave function
is given by:
%
%
\begin{equation}
\label{eq:wf}
	\Psi=\frac{1}{\sqrt{3}}\left( ddu \chi^{\lambda 3} + {\rm
	permutation}\right)\phi ,
\end{equation}
with
%
%
\begin{eqnarray}
\chi^{\lambda3}_\uparrow&=&{1\over\sqrt 6}(\downarrow\uparrow\uparrow+
\uparrow\downarrow\uparrow-2\uparrow\uparrow\downarrow),\nonumber\\
\chi^{\lambda3}_\downarrow&=&{1\over\sqrt 6}(2\downarrow\downarrow\uparrow-
\downarrow\uparrow\downarrow-\uparrow\downarrow\downarrow)\;.
\end{eqnarray}
Since the wave function $\Psi$ must be an eigenfunction of $j^2$ ($j$
being the total spin of the neutron) and the longitudinal component
$j_3$, the spins $\uparrow$ and $\downarrow$ have to be rotated by the
Melosh transform.$^{6,12}$ The $S$-state orbital function
$\phi (M_0)$ is chosen to be
%
%
\begin{equation}
\label{eq:pol}
	\phi(M_0)=\frac{N}{(M_0^2+\alpha^2)^n} ,
\end{equation}
with $\alpha $ and $n$ being phenomenological parameters and $N$ being
the normalization given by:
%
%
\begin{equation}
	\frac{N_c}{(2\pi)^6}\int d^3k d^3K \phi^2 =1 .
\end{equation}
The form factors are calculated by inserting Eq.~(\ref{eq:wf})
into Eq.~(\ref{eq:me}). The result is rather lengthy and the explicit
expressions are given in Ref.~8. The exponent $n$ is fitted
to the proton form factor $G_{Mp}$ giving $n=3.5$. The constituent quark
mass $m$ and the length scale parameter $\alpha $ are fitted to the
proton magnetic moment and the weak neutron decay, which results in
$m=0.263$~GeV and $\alpha =0.607$~GeV.

\begin{figure}[htbp]
\vspace{3.2in}
\ffcaption{The magnetic form factor of the neutron compared with the dipol
fit, $G_{Mn}/\mu_nG_D$. The experimental data are taken from
Ref.~1 with statistical and systematical errors.
Solid line, our calculation with pole type wave functions; dashed line,
our calculation with a harmonic oscillator type wave function;
dash-dotted line, VMD model from H\"{o}hler;$^3$
dash-double-dotted line, Gari-Kr\"{u}mpelmann model;$^5$
dotted line, QCD sum rule prediction by Radyushkin.$^{7}$}
\end{figure}

\begin{figure}[htbp]
\vspace{3.2in}
\ffcaption{The electric form factor of the neutron compared with the dipol
fit, $G_{En}^2/G_D^2$. The data and curves are marked the same as in
Fig.~1. }
\end{figure}

Figures~1 and 2 show the magnetic and electric
form factors of the neutron respectively. The figures give the deviation
from the dipol fit $G_D=(1+Q^2/M_V^2)^{-2}$ with $M_V=0.84$~GeV. Only
experimental data from SLAC~NE11~$^1$ are given since previous
data do not distinguish between the various theoretical predictions. The
present calculation (solid curves) is in very good agreement for both form
factors. There is only a slight deviation for the magnetic form factor
around 2~GeV$^2$. To show that the specific form of the wave function in
Eq.~(\ref{eq:pol}) is essential for the result we compare the result with
the commonly used exponential wave function $\phi
(M_0)=N\exp{(-M_0^2/2\alpha^2)}$. We also fixed the parameters by fitting
other electroweak nucleon properties,$^9$ and get $m=0.267$~GeV
and $\alpha=0.56$~GeV. The dashed line shows a rapid decrease for $G_{Mn}$
at already 1~GeV$^2$, which indicates that the exponential wave function is
not useful at that energy range. In the nonrelativistic limit, $\alpha/m
\to 0$, the form factors fall far below the dipol fit for any reasonable
value of $\alpha$ and $m$.\fnm{a}\fnt{a}{For the wave function in
Eq.~(\ref{eq:pol}) we get
$G_{Mn}=\mu_n (1+2Q^2/(\alpha^2+9m^2))^{-3.5}$ as the nonrelativistic
limit for low $Q^2$ and
$G_{Mn}=\mu_n (\sqrt{2} Q/3m)^{1/2}
(2Q^2/(\alpha^2+9m^2))^{-3.5}$ for high $Q^2$.} The relativistic treatment is
therefore important, which is a fact also observed for the pion.$^{17}$
The VMD model (dash-dotted curves) from
H\"{o}hler~$^3$ agrees with the $G_{En}$ data, but overestimates
$G_{Mn}$. The model from Gari and Kr\"{u}mpelmann $^5$
(dash-double-dotted curves) predicts $F_{1n}=0$. It is therefore in very
poor agreement with $G_{En}$, and in addition underestimates $G_{Mn}$. The
QCD sum rule predictions from Radyushkin~$^7$ (dotted curves)
agrees for $G_{En}$ and underestimates $G_{Mn}$, approaching $G_{Mn}$ for
high $Q^2$. The QCD sum rule is not valid in the infrared region
$Q^2<1$~GeV$^2$ due to singular power corrections at $Q=0$.\fnm{b}\fnt{b}
{A new method for QCD sum rules in the infrared region $0<Q^2<1$~GeV$^2$
is described in Ref.~18.}

We conclude that the precise measurement of form factors at intermediate
energies gives valuable constraints on theoretical models. We showed
that a model, that is in excellent agreement with the electroweak
properties of the baryon octet, gives a parameter free prediction of
the nucleon form factors, which is in good agreement with recent
experimental data. The relativistic features of the model and the
specific form of the wave function are essential for the good
result.

\nonumsection{Acknowledgements}
\noindent
This work was supported in part by the Schweizerischer Nationalfonds and
in part by the Department of Energy, contract DE-AC03-76SF00515.

\nonumsection{References}

\end{document}